\documentclass{article}
\usepackage{spconf,amsmath,graphicx}
\usepackage{epsfig}
\usepackage{algorithmic}
\usepackage{algorithm}
\usepackage{amssymb}
\usepackage[tight,footnotesize]{subfigure}
\usepackage{array}
\usepackage{color}

\DeclareGraphicsExtensions{.eps}

\newcolumntype{C}[1]{>{\raggedright\let\newline\\\arraybackslash}p{#1}}

\def\cosp {\ell} 

\def\spark{\mathrm{Spark}}

\newtheorem{thm}{Theorem}[section]

\newtheorem{prop}[thm]{Proposition}
\newtheorem{defn}[thm]{Definition}

\newcommand{\norm}[1]{\left\Vert#1\right\Vert}
\newcommand{\abs}[1]{\left\vert#1\right\vert}
\newcommand\vect[1]{{\bf#1}}
\newcommand\matr[1]{{\bf#1}}

\newcommand\alphabf{{\boldsymbol{\alpha}}}
\newcommand{\argmin}{\operatornamewithlimits{argmin}}

\newcommand{\Real}{\mathbb R}
\newcommand\RR[1]{\mathbb{R}^{#1}}

\DeclareMathOperator{\range}{range}

\title{Can We Allow Linear Dependencies in the Dictionary \\in the Sparse Synthesis Framework?}

\name{Raja~Giryes and Michael Elad\thanks{R. Giryes is grateful to the Azrieli Foundation for the award of
an Azrieli Fellowship. This research was partly supported by the ERC Advanced grant agreement no. 320649.
}}
\address{Department of Computer Science,
Technion - IIT 32000, Haifa, ISRAEL.}

\begin{document}

\maketitle

\begin{abstract}
Signal recovery from
a given set of linear measurements using a sparsity prior
has been a major subject of research in recent
years. In this model, the signal is assumed to have
a sparse representation under a given dictionary.
Most of the work dealing with this subject has focused on the reconstruction of the signal's
representation as the means for recovering the signal itself.
This approach forced the dictionary to be of
low coherence and with no linear dependencies between its columns.
Recently, a series of contributions that focus on signal recovery using the analysis model
find that linear dependencies in the analysis dictionary
are in fact permitted and beneficial.
In this paper we show theoretically that the same holds also for signal recovery
in the synthesis case for the $\ell_0$-synthesis minimization problem.
In addition, we demonstrate empirically the relevance of our conclusions
for recovering the signal using an $\ell_1$-relaxation.
\end{abstract}


\section{Introduction}
\label{sec:intro}

Recovering a signal from a noisy set of
linear measurements appears in many problems
such as compressed sensing, image deblurring and
super-resolution. In the basic setup an unknown signal
$\vect{x}_0\in \RR{d}$ passes through a given linear transformation $\matr{M} \in \RR{m\times d}$;
an additive noise $\vect{e} \in \RR{m}$ contaminates the outcome,
providing a set of linear measurements
\begin{eqnarray}
\label{eq:meas}
\vect{y} = \matr{M}\vect{x}_0 + \vect{e}.
\end{eqnarray}
The noise $\vect{e}$ can be assumed to be either an adversarial
unknown bounded noise such that $\norm{\vect{e}}_2 \le \epsilon$
or a random noise with a given distribution, e.g., zero-mean white Gaussian noise.
In this paper we focus on the first case.
The measurement matrix $\matr{M}$ is usually either rectangular with $m< d$ or an ill-posed matrix with $m=d$.
Thus, the information we have in the measurement vector $\vect{y}$ is not enough for recovering
$\vect{x}$, and additional prior knowledge is needed.

The priors considered in this work are sparsity based ones \cite{Bruckstein09From}.
These include two main paradigms for modeling the signal -- the synthesis and the analysis schemes \cite{elad07Analysis}. In the synthesis framework the signal $\vect{x}$ is assumed to
have a sparse representation $\alphabf \in \RR{n}$ under a given dictionary $\matr{D} \in \RR{d \times n}$,
\begin{eqnarray}
\vect{x}_0 = \matr{D}\alphabf_0.
\end{eqnarray}
Few words on our terminology: A vector $\alphabf$ is said to be $k$-sparse if $\norm{\alphabf}_0 \le k$,
where $\norm{\cdot}_0$ is the $\ell_0$ semi-norm that counts the number of its non-zero elements, and
a signal $\vect{x}$ is said to have a $k$-sparse representation if there exists a $k$-sparse $\alphabf$ such that $\vect{x} = \matr{D}\alphabf$.
The columns of $\matr{D}$ are referred to as atoms, the non-zero locations of a sparse vector $\alphabf$ as the support $T$,
the size of this support is $\abs{T}$,
and the sub-matrix of $\matr{D}$ with the atoms corresponding to $T$ as $\matr{D}_T$.
Given that $\vect{x}_0$ has a $k$-sparse representation, we can recover it from its measurement $\vect{y}$
by looking for the sparsest vector $\alphabf$ that satisfies $\norm{\vect{y}-\matr{M}\matr{D}\alphabf}_2\le\epsilon$
(In the noiseless case, $\epsilon=0$).

The analysis framework models the signal differently. Given a matrix $\matr{\Omega} \in \RR{p \times d}$ -- the analysis dictionary --
we are interested in the multiplication $\matr{\Omega}\vect{x}$, referred to as the analysis representation.
A signal $\vect{x}$ is said to be $\cosp$-cosparse if
$\norm{\matr{\Omega}\vect{x}}_0 \le p$-$\ell$, i.e., $\matr{\Omega}\vect{x}$ has $\ell$ zero elements.
In this case we recover the signal by looking for the most cosparse vector $\vect{x}$, i.e., the one with the largest number of zeros in $\matr{\Omega}\vect{x}$
that satisfies $\norm{\vect{y} - \matr{M}\vect{x}}_2 \le \epsilon$.
More details can be found in \cite{Nam12Cosparse,Giryes12Greedy}.

Unlike the analysis model which works in the signal domain and thus recovers it directly,
the synthesis one works in the representation domain and thus recovers the signal
indirectly by recovering first its representation.
Because of this, most of the work that studies theoretical guarantees for the synthesis framework
give bounds for the representation's reconstruction,
or deal only with the case that $\matr{D}$ is unitary, for which the two problems coincide \cite{Donoho03Optimal, Donoho06OnTheStability, Candes06Near,Candes05Decoding,Needell09CoSaMP,Dai09Subspace,Blumensath09Iterative,Foucart11Hard}.

Results for more general dictionaries also exist \cite{Rauhut08Compressed,Tropp04Greed}.
However, they require the dictionary to be highly incoherent
and with no linear dependencies between small number of its columns.
This requirement limits the types of dictionaries that can be used to model the signal.
While this constraint is necessary for recovering the signal's representation,
it is not clear that it is still required when our target is the signal.
The goal of this paper is to answer this very question. We aim to show that highly-coherent and even linearly dependent atoms in $\bold D$ still enable a reliable recovery of the signal from its measurements.

Recently, a series of works \cite{Candes11Compressed,Nam12Cosparse,Giryes12Greedy,Rubinstein12Cosparse,Peleg12Performance}
showed that in the analysis framework, where we aim at recovering the signal directly,
linear dependencies can be allowed in the (analysis) dictionary.
These correlations were even favored as they were shown to improve the recovery performance of the different techniques analyzed in these works.

Since linear dependencies are shown to give profit in the analysis framework,
it is conjectured that the requirement for an incoherent dictionary
for the signal recovery in the synthesis framework is unnecessary as well.
A clue for this very property is given in \cite{Candes11Compressed}, where the reconstruction conditions are presented in terms of $\matr{D}$-RIP which is a property of the measurement matrix $\matr{M}$ for the synthesis model. However, the results in \cite{Candes11Compressed} are derived for signals from the analysis model, thus leaving our question unresolved as of yet.

In this work we focus on the synthesis model and ask: Is it possible
to have a recovery guarantee for the synthesis model for a dictionary
exhibiting linear dependencies within small groups of its columns?
In order to answer this question, we study the performance of the ideal
$\ell_0$-minimization problem
\begin{eqnarray}
\label{eq:l0_min}
\hat\alphabf = \argmin_\alphabf \norm{\alphabf}_0 & s.t. & \norm{\vect{y} - \matr{M}\matr{D}\alphabf}_2 \le \epsilon.
\end{eqnarray}
This is the core approach for recovering a signal that is known to have the synthesis sparsity prior.
The reconstructed signal in this technique is simply $\hat{\vect{x}} = \matr{D}\hat{\alphabf}$.
In the noiseless case, \eqref{eq:l0_min} turns to be simply
\begin{eqnarray}
\label{eq:l0_min_noiseless}
\hat\alphabf = \argmin_\alphabf \norm{\alphabf}_0 & s.t. & \vect{y} = \matr{M}\matr{D}\alphabf.
\end{eqnarray}

In this work we study the recovery performance of \eqref{eq:l0_min} and \eqref{eq:l0_min_noiseless}. We first provide uniqueness conditions for the signal recovery in the noiseless case. Then we present stable reconstruction conditions for the noisy case where the noise is adversarial. This result is a particular case of the result
presented in \cite{Blumensath09Sampling,Lu08Theory} in which a more general form of the $\matr{D}$-RIP property was proposed
giving stable recovery guarantee for signals that come from a general union of subspaces model.

The uniqueness and stability conditions that we present are generalization of previous results \cite{Elad02generalized, Donoho03Optimal, Donoho06OnTheStability,
Candes06Near,Candes05Decoding,Candes06Modern} that assume $\matr{D}$ to be incoherent.
The contribution of this work is in providing reconstruction guarantees in the signal domain that do not pose
incoherence requirement on the dictionary.

The $\ell_0$-minimization problem is known to be NP-hard \cite{NP-Hard} and several approximation techniques
have been suggested to it \cite{Elad02generalized, Donoho03Optimal, Needell09CoSaMP, Dai09Subspace, Blumensath09Iterative, Foucart11Hard, Candes07Dantzig, MallatZhang93}.
A future work should extend the existing representation recovery guarantees to the signal case.
First steps in this direction have been taken for the CoSaMP algorithm in \cite{Davenport12Signal}.
However, the theoretical study of this algorithm is impaired by the assumed existence of a near-optimal projection scheme in the signal domain, an assumption similar to the one posed in \cite{Giryes12Greedy}. The existence of such a projection in the general case is still an open question.

In addition to the theoretical study of the $\ell_0$-problem, in this paper we show empirically that for the
$\ell_1$-synthesis minimization problem, a relaxed version of \eqref{eq:l0_min} that replaces the $\ell_0$-semi-norm with an $\ell_1$-norm,
signal recovery is at hand in the cases where $\matr{D}$ is highly coherent and the representation recovery is impossible.
Similar experiments has been performed in \cite{Davenport12Signal} for the CoSaMP algorithm.

The organization of this paper is as follows.
In Section~\ref{sec:back} we survey existing uniqueness and stability guarantees for the representation reconstruction using the $\ell_0$-minimization problem.
In Section~\ref{sec:signal} the previously presented results
are extended for the signal reconstruction case, allowing general coherent dictionaries.
Empirical reconstruction results for the $\ell_1$-minimization problem are presented in Section~\ref{sec:exp}, comparing
the representation and signal recovery in the case where $\matr{D}$ is highly coherent.
In Section~\ref{sec:conc} the work is summarized.

\section{Background - Guarantees for the Representation Recovery}
\label{sec:back}

Two important properties that were explored for the $\ell_0$-problem
are the uniqueness of its solution and the stability of the solution under bounded adversarial noise.
In this section we survey the existing results for representation recovery.
We present the results with no proofs, as they are already contained in those of Section~\ref{sec:signal}
for signal recovery.

\subsection{Uniqueness}

Given a signal's representation $\alphabf_0$ with a cardinality $k$ ($\norm{\vect{\alphabf_0}}_0 \le k$) we are interested to know whether it is the unique solution of \eqref{eq:l0_min_noiseless}.
In other words, whether there exists another representation
$\alphabf_1 \ne \alphabf_0$ with a cardinality at most $k$ such that $\matr{M}\matr{D}\alphabf_0 = \matr{M}\matr{D}\alphabf_1$.
An answer for this question was given in \cite{Donoho03Optimal} using the definition of the Spark
of a matrix:
\begin{defn}[Definition 1 in \cite{Donoho03Optimal}]
Given a matrix $\matr{A}$ we define $\spark(\matr{A})$ as
the smallest possible number of
columns from $\matr{A}$ that are linearly dependent.
\end{defn}
This definition provides us with a sharp uniqueness condition for the representation reconstruction:
\begin{thm}[Corollary 3 in \cite{Donoho03Optimal}]
Let $\vect{y}=\matr{M}\matr{D}\alphabf_0$.
If $\norm{\alphabf_0}_0<\spark(\matr{M}\matr{D})/2$ then $\alphabf_0$ is the unique solution of \eqref{eq:l0_min_noiseless}.
\end{thm}
Though this condition is sharp for finding the representation, it is not sharp at all, in terms of finding the signal itself.
This can be demonstrated using the following simple example. Let us assume that $\matr{D}=[\vect{z},\vect{z},\dots,\vect{z}]$,
a dictionary with columns that are a duplicate of the same atom $\vect{z}$. Clearly, the signal $\vect{x}$=$\vect{z}$ can be represented by any of the atoms of $\matr{D}$, which means that it has $n$ different sparse representations, each
with cardinality $1$. Thus, for any measurement matrix $\matr{M}$ there is no unique solution to \eqref{eq:l0_min_noiseless}.
Indeed, we have $\spark(\matr{M}\matr{D})=2$ and the uniqueness condition collapses to the trivial requirement $\norm{\alphabf_0}_0 =0$.
However, it is clear that if our goal is to recover the signal $\vect{x}$ (i.e. $\matr{D}\alphabf$) and not its representation $\alphabf$,
then we can certainly have a uniqueness, as all the possible solutions to \eqref{eq:l0_min_noiseless} lead to the same signal.
Thus, we conclude that for the task of estimating the signal itself,
the existing condition is not sharp and a better one should be explored.

\subsection{Stability}

In the case where noise exists in the measurement, the uniqueness of the solution
is no longer the question since we cannot recover the original signal exactly.
Instead, we ask whether the reconstruction is stable, i.e., whether the $\ell_2$-distance between
the estimated signal's representation and the original one is proportional to the noise power $\epsilon$.
In order to establish that, we use the restricted isometry property (RIP) \cite{Candes06Near, Candes05Decoding}. 
The RIP can be seen as an extension of the Spark, which allows noisy case analysis.
\begin{defn}
A matrix $\matr{A}$ satisfies the RIP
condition with parameter $\delta_k$ if it is the smallest value that
satisfies
\begin{equation}
(1-\delta_k) \norm{\alphabf}_2^2 \le \norm{\matr{A}\alphabf}_2^2 \le
(1+\delta_k) \norm{\alphabf}_2^2
\end{equation}
for any $k$-sparse vector $\alphabf$.
\end{defn}
The connection between the RIP and the Spark is the following.
Given a matrix $\matr{A}$, $k < \spark(\matr{A})$ if $\delta_k<1$ \cite{Candes05Decoding}. Having the RIP it is straight forward to have a stability condition for the representation reconstruction using \eqref{eq:l0_min}.
\begin{thm}
Let $\matr{y}= \matr{M}\matr{D}\alphabf_0 + \vect{e}$ where $\norm{\vect{e}}_2 \le \epsilon$, $\matr{M}\matr{D}$ satisfies the RIP condition with $\delta_{2k}$ and $\norm{\alphabf_0}_0 \le k$.
If $\delta_{2k}<1$ then the solution $\hat\alphabf$ of \eqref{eq:l0_min} is stable.
More specifically,
\begin{eqnarray}
\norm{\alphabf_0 - \hat\alphabf}_2 \le  \frac{2\epsilon}{\sqrt{1-\delta_{2k}}}.
\end{eqnarray}
\end{thm}
The same problem we had with the dictionary $\matr{D}= [\vect{z},\vect{z},\dots,\vect{z}]$ in the uniqueness case
repeats also here because 
$\delta_{2k}<1$ implies $2k<\spark(\matr{MD})$.

\section{Guarantees for Signal Recovery}
\label{sec:signal}

As mentioned before, though the Spark and RIP conditions for $\matr{MD}$ are sharp for the representation recovery, they are not designed for the signal recovery. In this section we use an extended Spark and RIP definitions that will serve better the signal recovery problem. The $\matr{D}$-RIP \cite{Candes11Compressed} is used for having stable recovery conditions for the signal reconstruction. In a similar way, we propose a $\matr{D}$-Spark property for the measurement matrix $\matr{M}$, introducing a new uniqueness condition for the signal recovery.
Note that the results in this section are generalization of the ones presented in the previous section for the signal's representation.
As a general guideline, by setting the measurement matrix to be $\matr{MD}$ and the dictionary to be the identity, the results of this section coincide with those of the previous one.

\subsection{Uniqueness for Signal Recovery}

As in the representation case, we are interested to know when we can guarantee that a signal $\vect{x}_0$ with a $k$-sparse representation under a matrix $\matr{D}$ is the unique solution of \eqref{eq:l0_min_noiseless}.
In other words, whether there exist another signal $\vect{x}_1 \ne \vect{x}_0$ with at most $k$-sparse representation
under $\matr{D}$ such that $\matr{M}\vect{x}_0 = \matr{M}\vect{x}_1$.
For this task we introduce the $\matr{D}$-Spark, an extension of the Spark definition.
\begin{defn}
Given a matrix $\matr{M}$ we define $\matr{D}$-$\spark(\matr{M})$ as
the smallest possible number of columns from $\matr{D}$, marked as $T$,
such that $\range(\matr{D}_T)\cap \mathbf{Null}(\matr{M}) \ne \{ 0 \}$.
\end{defn}
In other words, for every set $T$ with size $\abs{T} < \matr{D}$-$\spark(\matr{M})$ we
have $\range(\matr{D}_T)\cap \mathbf{Null}(\matr{M}) = \{ 0 \}$,
implying that for any vector $\vect{v}\in \Real^{\abs{T}}$, $\matr{MD}_T\vect{v}=0$ if and only if $\matr{D}_T\vect{v}=0$. 
Note that this definition coincides with the one of the Spark for $\matr{D}$=$\matr{I}$.
This can be observed by noticing that in this case $\matr{D}_T$=$\matr{I}_T$ and thus $\matr{MD}_T$
simply chooses $T$ columns from $\matr{M}$.
Thus, the above translates to the requirement that there is no subset of $\abs{T}$ columns in $\matr{M}$ that are linearly dependent.

Having the $\matr{D}$-Spark definition, we propose a uniqueness condition for the signal recovery.
\begin{thm}
Let $\vect{y} = \matr{M}\matr{x}_0$ where $\vect{x}_0$ has a $k$-sparse representation $\alphabf_0$ under $\matr{D}$.
If $k< \matr{D}$-$\spark(\matr{M})/2$ then $\vect{x}_0 = \matr{D}\hat{\alphabf}$ for
$\hat{\alphabf}$ the minimizer of \eqref{eq:l0_min_noiseless}, implying a perfect recovery.
\end{thm}
{\em Proof:}
Let us assume the contrary, i.e., there exists a minimizer, $\hat\alphabf_1$ for \eqref{eq:l0_min_noiseless} such that $\matr{D}\hat\alphabf_1 \ne \matr{D}\alphabf_0$.
Let us denote the support sets of $\alphabf_0$ and $\hat\alphabf_1$ by $T_0$ and $T_1$ respectively.
Since $\alphabf_0$ is a feasible solution to \eqref{eq:l0_min_noiseless} and $\hat\alphabf_1$ is a minimizer, we have that $\abs{T_1} \le \abs{T_0}\le k$.
Thus, by definition $\matr{D}\hat\alphabf_1 - \matr{D}\alphabf_0 \in \range(D_{T_1\cup T_0})$ where $\abs{T_1 \cup T_0} \le 2k$.
Noticing that $\matr{M}\matr{D}\hat\alphabf_1 = \matr{M}\matr{D}\alphabf_0$, by the constraint of  \eqref{eq:l0_min_noiseless}, we have $\matr{D}\hat\alphabf_1 - \matr{D}\alphabf_0 \in \mathbf{Null}(\matr{M})$.
This contradicts the assumption $k< \matr{D}$-$\spark(\matr{M})/2$ because we get that
$\matr{D}\hat\alphabf_1 - \matr{D}\alphabf_0 \in \range(\matr{D}_{T_1\cup T_0})\cap \mathbf{Null}(\matr{M}) \ne \{ 0 \}$, which means that $\matr{D}$-$\spark(\matr{M}) < \abs{T_1\cup T_0} \le 2k$.
\hfill $\Box$
\smallskip

Unlike the uniqueness condition of the regular Spark, the one of the $\matr{D}$-Spark allows linear dependencies within the dictionary $\matr{D}$. Looking at the example from the previous section, for the dictionary $\matr{D} = [\vect{z},\vect{z},\dots,\vect{z}]$, $\range(\matr{D}_T) = \{\beta\vect{z}, \beta \in \Real\}$ for any non-empty support $T$. Thus, the uniqueness condition turns out to be
$\vect{z} \not\in \mathbf{Null}(\matr{M})$. This means that the family of matrices $\matr{M}$ that guarantee uniqueness are the ones that have at least one row non-orthogonal to $\vect{z}$. This is far stronger compared to earlier condition as discussed in Section~\ref{sec:back}.

\subsection{Stability for Signal Recovery}

Moving to the noisy case, we seek for a generalization of the RIP that provides us with guarantees
for the signal recovery. For this task we use the $\matr{D}$-RIP, as introduced in \cite{Candes11Compressed}.
\begin{defn}
A matrix $\matr{M}$ satisfies the $\matr{D}$-RIP
condition with parameter $\delta_k^{\matr{D}}$ if it is the smallest value that
satisfies
\begin{equation}
(1-\delta_k^{\matr{D}}) \norm{\vect{x}}_2^2 \le \norm{\matr{M}\vect{x}}_2^2 \le
(1+\delta_k^{\matr{D}}) \norm{\vect{x}}_2^2
\end{equation}
for any vector $\vect{x}$ that has a $k$-sparse representation under $\matr{D}$,
or equivalently for any $\vect{x} \in \range(\matr{D}_T)$, where $T$ is such that $\abs{T}\le k$.
\end{defn}
As in the representation case, a connection between the $\matr{D}$-RIP and the $\matr{D}$-Spark can be established.
\begin{prop}
Given a matrix $\matr{M}$ and sparsity $k$, if $\delta_k^{\matr{D}}<1$ then $k < \matr{D}$-$\spark(\matr{M})$ .
\end{prop}
 {\em Proof:}
Requiring $\delta_k^{\matr{D}} <1$ implies
that for any vector $\vect{x} \in \range(\matr{D}_T)$ such that $\abs{T}\le k$ it holds that $\norm{\matr{M}\vect{x}}_2 \ge (1-\delta_k^{\matr{D}})\norm{\vect{x}}_2>0$, 
hence $\matr{M}\vect{x} \ne 0$ . The last is equivalent to requiring $\mathbf{Null}(\matr{M})\cap \range(\matr{D}_T) = \{ 0 \}$ for any support set $T$ such that $\abs{T}\le k$, which is exactly equivalent to $\matr{D}$-$\spark(\matr{M}) > k$.
\hfill $\Box$
\smallskip

Having the definition of the $\matr{D}$-RIP we present a stability guarantee for the signal recovery that appears in \cite{Blumensath09Sampling,Lu08Theory}.
\begin{thm}
Let $\matr{y}= \matr{M}\matr{x}_0 + \vect{e}$ where $\norm{\vect{e}}_2 \le \epsilon$, $\vect{x}_0$ has a $k$-sparse representation $\alphabf_0$ under $\matr{D}$, and $\matr{M}$ satisfies the $\matr{D}$-RIP condition with $\delta_{2k}^{\matr{D}}$.
If $\delta_{2k}^{\matr{D}}<1$ then recovering the signal using \eqref{eq:l0_min}, where the recovered signal is $\hat{\vect{x}} = \matr{D}\hat\alphabf$, is stable: 
\begin{eqnarray}
\norm{\vect{x}_0 - \hat{\vect{x}}}_2 \le  \frac{2\epsilon}{\sqrt{1-\delta_{2k}^{\matr{D}}}}.
\end{eqnarray}
\end{thm}
 {\em Proof:}
Since $\alphabf_0$, the representation of $\vect{x}_0$, is a feasible solution to \eqref{eq:l0_min}
we have that $\norm{\hat\alphabf}_0 \le \norm{\alphabf_0}_0 \le k$. Thus, $\vect{x}_0 - \hat{\vect{x}}$ is a signal that has a $2k$-sparse representation.
According to the constraint in \eqref{eq:l0_min} we also have $\norm{\matr{y}-\matr{M}\hat{\vect{x}}}_2 \le \epsilon$.
Using the $\matr{D}$-RIP, the triangle inequality, and the fact that $\norm{\matr{y}-\matr{M}{\vect{x}_0}}_2 \le \epsilon$ as well, we get
\begin{eqnarray}
&& \hspace{-0.25in}\norm{\vect{x}_0 - \hat{\vect{x}}}_2 \le \frac{1}{\sqrt{1-\delta_{2k}^{\matr{D}}}}\norm{\matr{M}(\vect{x}_0 - \hat{\vect{x}})}_2 \\ \nonumber && \hspace{-0.25in}
\le \frac{1}{\sqrt{1-\delta_{2k}^{\matr{D}}}}\left(\norm{\vect{y} - \matr{M}\vect{x}_0}_2 + \norm{\vect{y} - \matr{M}\hat{\vect{x}}}_2\right) \le \frac{2\epsilon}{\sqrt{1-\delta_{2k}^{\matr{D}}}},
\end{eqnarray}
which is the stated result.
\hfill $\Box$

\begin{figure}[htb]
\begin{minipage}[b]{.48\linewidth}
  \centering
  \centerline{\includegraphics[width=4.0cm]{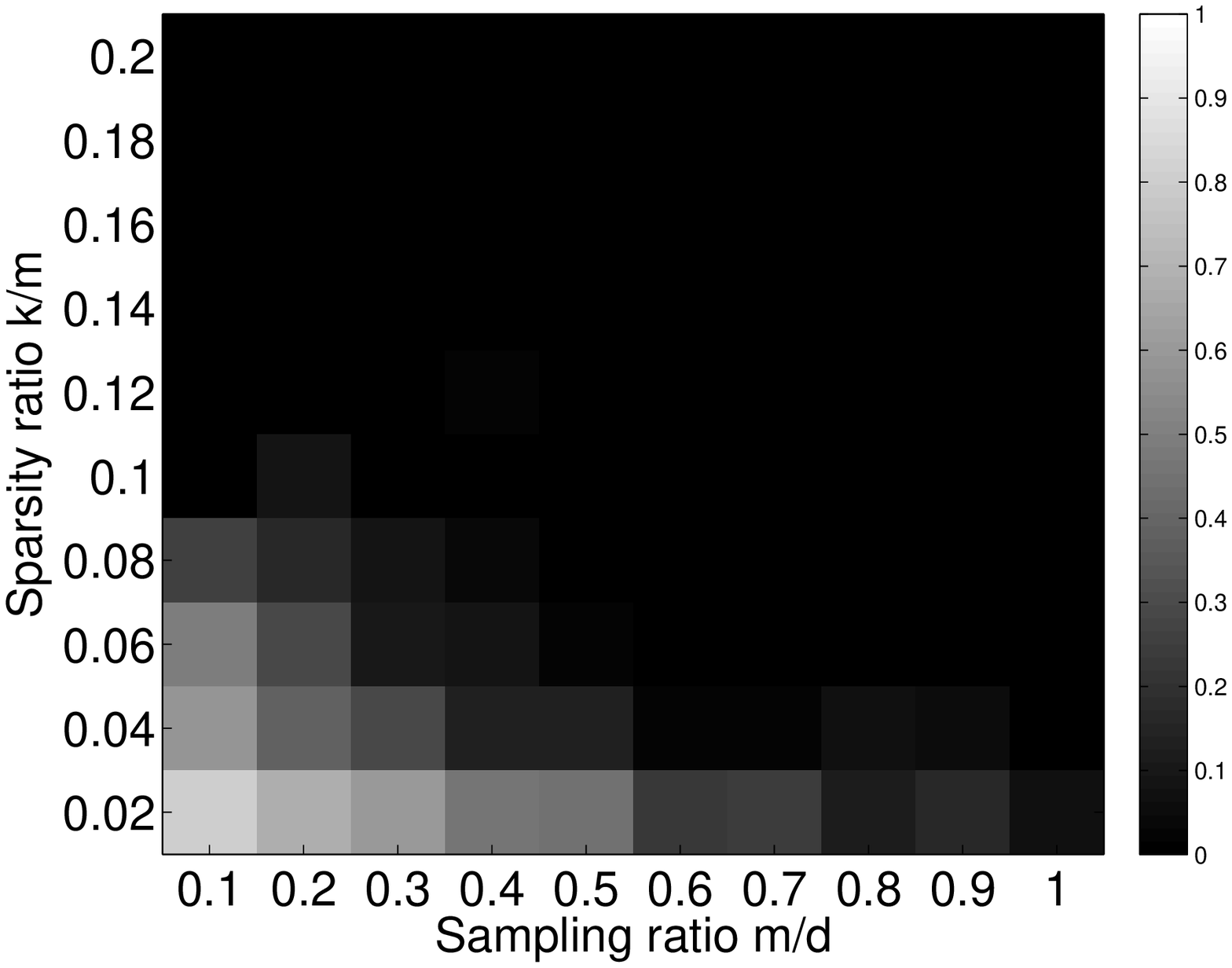}}
\end{minipage}
\hfill
\begin{minipage}[b]{.48\linewidth}
  \centering
  \centerline{\includegraphics[width=4.0cm]{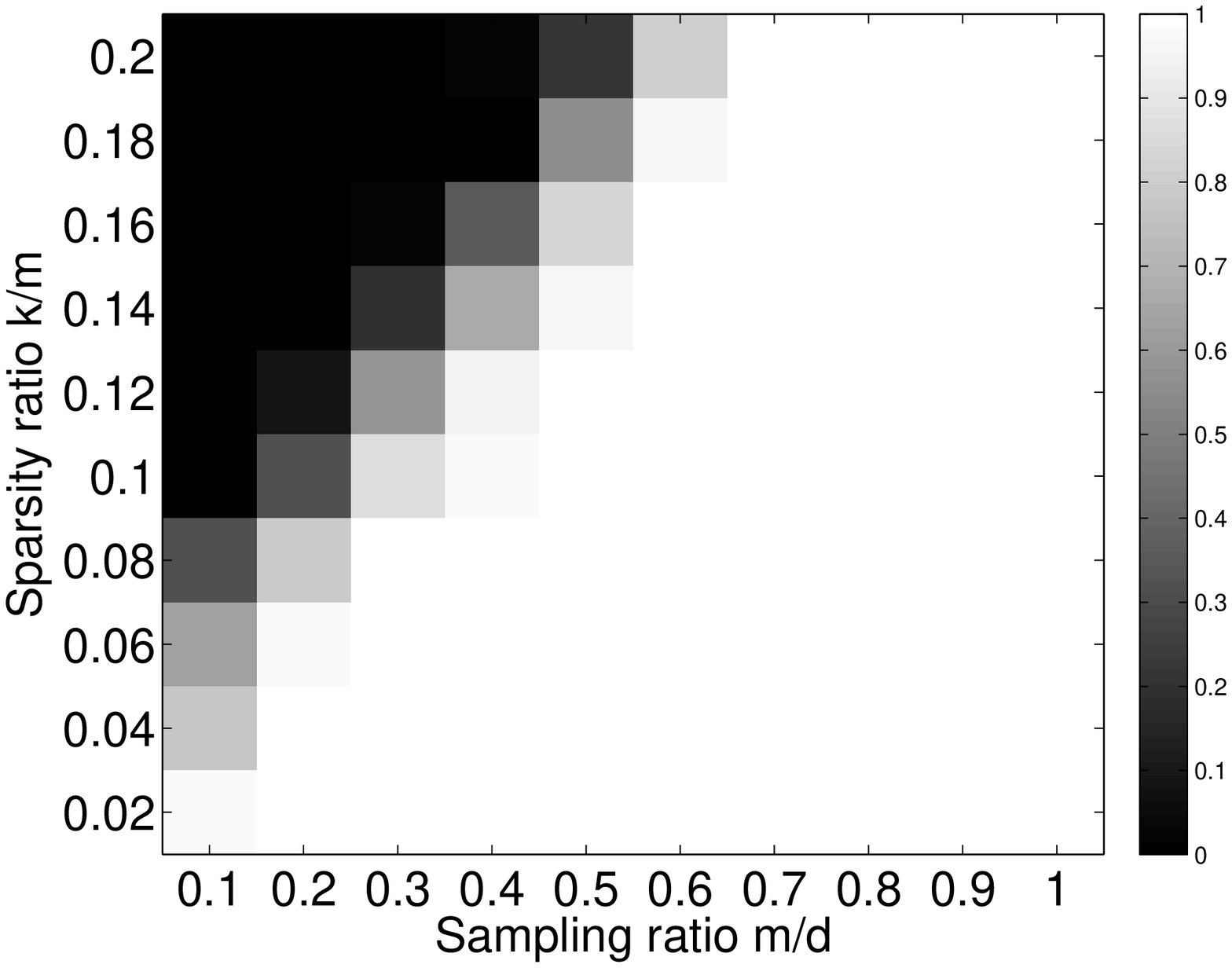}}
\end{minipage}
\caption{Representation recovery rate (left) and signal recovery rate (right). Color attribute: fraction of realizations in which $\ell_1$ minimization achieves a perfect recovery.}
\label{fig:recovery error}
\vspace{-0.15in}
\end{figure}

\section{Practical Reconstruction}
\label{sec:exp}
With the above observations, we turn to check the recovery performance of the $\ell_1$-minimization problem -- a relaxed version
that replaces the $\ell_0$ with $\ell_1$-norm.
In our experiment we consider the noiseless scenario, and since solving \eqref{eq:l0_min_noiseless} is NP-hard \cite{NP-Hard}, the $\ell_1$ relaxation is inevitable when dealing with practical problems. We perform a simple synthetic test
for signals that are sparse under a dictionary which is highly coherent and with linear dependencies between its columns.
We generate a dictionary $\matr{D}=[\matr{D}_1,\matr{D}_2]$ where $\matr{D}_1, \matr{D}_2\in \RR{d \times d}$, d=1000, $\matr{D}_1$ contains sparse columns with $2$ non-zero entries which are $1$ or $-1$ with probability $0.5$ and $\matr{D_2}$ contains columns which are linear combinations of random $3$ columns from $\matr{D}_1$ with random zero-mean white Gaussian weights. Each entry of the measurement matrix $\matr{M}\in \RR{m \times d}$ is distributed according to a normal Gaussian distribution,
where $m$=$\lfloor \gamma d \rfloor$ and $\gamma$ is the sampling rate -- a value in the range $(0,1]$.
We set $k$ to be $\lfloor \rho m \rfloor$ ($\rho \ll 1$) and measure the recovery rate of the representation $\alphabf$ and the signal $\vect{x}$ for various values of $\gamma \in \{0.1, 0.2, \dots, 1 \}$ and $\rho \in \{0.02,0.04,\dots, 0.2 \}$.

Figure~\ref{fig:recovery error} presents the recovery performance over $100$ realizations per each parameter setting.
Similar to the uniqueness results for the representation, the reconstruction fails almost always since $\matr{D}$ contains many linear dependencies (note that in our case $\spark(\matr{MD})=4$).
However, though we fail in reconstructing the representations, in most cases the recovery of the signal succeeds.
Note that even in the case where $m$=$d$ ($\gamma$=$1$), we get a very low recovery rate for the representations and this is due to the non-uniqueness of any representation with cardinality beyond $1$. The recovery rate on the bottom left part of the representation diagram is better than the other parts because for very small values of $k$ the chance to choose the wrong representation decreases significantly.


\section{Conclusion}
\label{sec:conc}
In this work we studied the $\ell_0$-synthesis problem for signal recovery in the case
of a general dictionary $\matr{D}$.
We have shown that in the case where $\matr{D}$ contains linear dependencies signal recovery is still theoretically plausible.
We derived theoretical conditions for uniqueness of the solution in the noiseless case and stability for the noisy case.
In addition, empirical reconstruction results were presented for the $\ell_1$-synthesis minimization problem, demonstrating
that for this problem signal recovery is achievable in the case where $\matr{D}$ is highly coherent.
These results motivate looking for theoretical guarantees for the signal recovery using the $\ell_1$-relaxation in the same
spirit of those developed in the representation case \cite{Elad02generalized,Donoho03Optimal,Candes06Near},
a topic that would be covered by our upcoming research.


\bibliographystyle{IEEEtran}
\bibliography{../UoS_theory,IEEEabrv}

\end{document}